\documentclass[aps,prl,twocolumn, showpacs, amsmath,longbibliography]{revtex4-2}  

\usepackage{bbm}
\usepackage{mathrsfs}
\usepackage{amsmath}
\usepackage{amsfonts}
\usepackage[colorlinks=true,citecolor=blue,anchorcolor=blue]{hyperref}
\usepackage{graphicx,epstopdf}
\usepackage{subfigure}
\usepackage{epsfig}
\usepackage{dcolumn}
\usepackage{bm}
\usepackage{color}
\usepackage{natbib}
\usepackage{amssymb}
\usepackage{xcolor}
\usepackage{braket}

\begin{document}

\bibliographystyle{apsrev4-2}

\title{Nonlinear Hall Effect in Antiferromagnetic Half-Heusler Materials}

\author{Cheng Chen$^{1,\dagger}$, Huaiqiang Wang$^{1,\dagger}$, Zhilong Yang$^{1}$ and Haijun Zhang$^{1,2,*}$}
\affiliation{
$^1$ National Laboratory of Solid State Microstructures, School of Physics, Nanjing University, Nanjing 210093, China\\
$^2$ Collaborative Innovation Center of Advanced Microstructures, Nanjing University, Nanjing 210093, China\\
}

\date{\today}


\begin{abstract}

It has recently been demonstrated that various topological states, including Dirac, Weyl, nodal-line, and triple-point semimetal phases, can emerge in antiferromagnetic (AFM) half-Heusler compounds. However, how to determine the AFM structure and to distinguish different topological phases from transport behaviors remains unknown. We show that, due to the presence of combined time-reversal and fractional translation symmetry, the recently proposed second-order nonlinear Hall effect can be used to characterize different topological phases with various AFM configurations. Guided by the symmetry analysis, we obtain the expressions of the Berry curvature dipole for different AFM configurations. Based on the effective model, we explicitly calculate the Berry curvature dipole, which is found to be vanishingly small for the triple-point semimetal phase, and large in the Weyl semimetal phase. Our results not only put forward an effective method for the identification of magnetic orders and topological phases in AFM half-Heusler materials, but also suggest these materials as a versatile platform for engineering the non-linear Hall effect.

\end{abstract}

\email{zhanghj@nju.edu.cn}


\maketitle



Since the discovery of the topological nature of quantum Hall effect \cite{thouless1982}, many topological states and topological materials, including topological insulators \cite{Hasan2010,Qi2011,koenig2007,Zhang2009,Yu2010,Chang2013}, topological semimetals \cite{Wan2011,xu2011,Wang2012,Wang2013,weng2015,Ruan2016a,Bradlyn2016,armitage2018}, topological superconductors \cite{fu2008,Qi2009,sun2016,zhang2018,fang2019,yuan2019}, topological magnetic axion insulator \cite{Mong2010,zhang2019,li2019,xu2019,liu2020,zhang2020cpl,wang2020dynamical}, and so on, have been found, which has significantly improved our understanding of symmetry and topology in condensed matter physics. Quasiparticle excitations near certain band crossing points, protected by specific space-group symmetries \cite{Wieder2016, Bradlyn2016}, can be considered as analogs of elementary particles, such as, Dirac and Weyl quasiparticles; these materials are known as Dirac and Weyl semimetals \cite{Wan2011, Wang2012}. As typical topological materials, half-Heusler compounds have attracted widespread attention. Both topological insulators and topological semimetals have been discovered in half-Heusler compounds, both with and without external perturbations such as strains and magnetic fields \cite{Lin2010,Chadov2010,Xiao2010a, AlSawai2010, Yan2014, Ruan2016a,  Hirschberger2016, Yang2017, Cano2017, Shekhar2018}. Intriguingly, it has recently been shown that in the presence of antiferromagnetic (AFM) order, rich magnetic topological phases can emerge in AFM half-Heusler materials, including AFM topological insulator, Dirac/Weyl semimetal, and nodal-line semimetal, and triple-point (TP) semimetal phases \cite{Mong2010,Suzuki2016,Yu2017,Shekhar2018}. This indicates AFM half-Heusler materials could be a versatile platform from which researchers can study magnetic topological states. Since magnetic topological states depend on a detailed AFM configuration, the problem of how to distinguish magnetic topological states in AFM half-Heusler materials turns out to be a significant issue.

Topologically nontrivial states usually manifest via unique transport behaviors, such as quantized Hall conductivity for the quantum Hall effect with broken time-reversal symmetry \cite{Klitzing1980, Chang2013}. Recently, a second-order nonlinear Hall effect (NLHE) was proposed in relation to noncentrosymmetric materials in the presence of the time-reversal symmetry \cite{Sodemann2015}, which essentially originates from the Berry curvature dipole (BCD) of electronic structures. Moreover, the NLHE has also been predicted in transition metal dichalcogenides \cite{Du2018, Zhang2018a, You2018} and Weyl semimetals \cite{Sodemann2015,Facio2018,Zhang2018c, Chen2019}. Shortly after this, the NLHE was experimentally observed \cite{Xu2018, Kang2018, Ma2019, Son2019}. As such, the NLHE offers an effective approach to the characterization of topological states. It is worth mentioning that the NLHE has been extended to include AFM materials \cite{Shao2020} with $\mathcal{S}=\mathcal{T}\tau_{1/2}$ symmetry, i.e., a combination of time-reversal symmetry $\mathcal{T}$ and a fractional translation $\tau_{1/2}$. Since the BCD is present even under $\mathcal{S}$ symmetry, it plays a similar role to the time-reversal symmetry $\mathcal{T}$. This therefore allows us to investigate different magnetic topological states in AFM half-Heusler materials with the $\mathcal{S}$ symmetry preserved, via the NLHE method. 

In this work, we study the NLHE of magnetic topological states in AFM half-Heusler materials. Firstly, based on the standard effective Luttinger Hamiltonian, we begin with the nonmagnetic parent states, and treat the AFM order as a perturbation in order to calculate electronic structures and phase diagrams under different AFM orders. Next, using symmetry analysis, we derive the independent, symmetry-permitted, nonvanishing components of the BCD tensor. Finally, we focus on the TP and Weyl semimetal phases to explicitly calculate the BCD. A prominent peak of BCD is found near the Weyl points (WPs), while no pronounced BCD features are found for TPs; our results are further verified by calculations based on low-energy effective Hamiltonians near these band crossing points.

\begin{figure}[htbp]
    \includegraphics[width =2.5in]{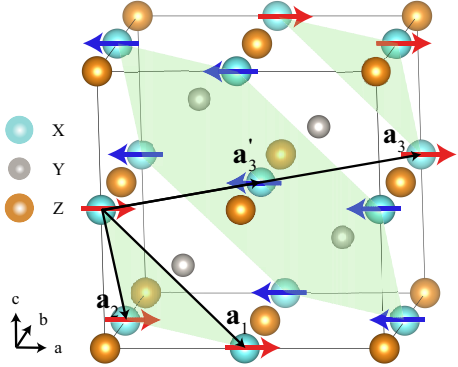}
    \caption{\label{fig:half-Heusler} Crystalline structure and magnetic order of half-Heusler materials $XYZ$. $X$ and $Z$ atoms form the NaCl-type substructure, and $Y$ and $Z$ atoms form the zinc-blende substructure. Magnetic moments are generated by the $X$ atoms, and the  magnetic moments are schematically shown by arrows. Note that magnetic moments align ferromagnetically in the layer perpendicular to the $[111]$ direction, and antiferromagnetically between two adjacent layers, in the typical G-type AFM. In the absence of the magnetic order, $\boldsymbol{a}_1$, $\boldsymbol{a}_2$, $\boldsymbol{a}_3^{\prime}$ constitute the lattice basis vectors. With respect to the AFM order, the lattice vector along the $\boldsymbol{a}_3^{\prime}$ direction is doubled . The new lattice basis vector are labeled by $\boldsymbol{a}_1$, $\boldsymbol{a}_2$, and $\boldsymbol{a}_3=2\boldsymbol{a}_3^{\prime}$.}
\end{figure}

\textit{Topological Phases.} Half-Heusler compounds labeled by $XYZ$ are a group of materials which have been extensively studied, particularly in relation to topological states \cite{Lin2010, Graf2011}. The compound's crystalline structure is shown in Fig.~\ref{fig:half-Heusler}, where $X$ and $Z$ atoms form the NaCl-type substructure, and $Y$ and $Z$ atoms form the zinc-blende substructure. In the absence of AFM order, the crystal space group and corresponding point group are $F_{\overline43m}$ and $T_d$ respectively. Here, $\boldsymbol{a}_1$, $\boldsymbol{a}_2$, and $\boldsymbol{a}_3^{\prime}$ are the original lattice basis vectors. With regard to the AFM order, the lattice vectors and unit cell are doubled along the $\boldsymbol{a}_3^{\prime}$ direction. The new lattice vectors are labeled by $\boldsymbol{a}_1$, $\boldsymbol{a}_2$, and $\boldsymbol{a}_3=2\boldsymbol{a}_3^{\prime}$. For AFM half-Heusler materials, such as GdPtBi, magnetic moments are generated primarily by the $X$ atoms, and align ferromagnetically in one layer which is perpendicular to the $[111]$ direction, and antiferromagnetically between two adjacent layers, in the G-type AFM order. 

Similarly to the zinc-blende semiconductors, such as HgTe, the electronic structure around the Fermi level of nonmagnetic half-Heusler materials is dominated by the $\Gamma_8$ bands \cite{Ruan2016a}. The $\Gamma_8$ bands form a fourfold degeneracy at the $\Gamma$ point, protected by time-reversal symmetry and the $T_d$ group. Along the $C_{3v}$ axis (e.g., the $[111]$ direction), the $\Gamma_8$ bands split into one doubly degenerate band, $\Lambda_6$, protected by $C_{3v}$ symmetry, and two nondegenerate bands, $\Lambda_{4,5}$, due to the inversion symmetry breaking of the $T_d$ group. This low-energy band structure is effectively described by the Luttinger Hamiltonian $H_0$ \cite{Zhang2018b} plus a linear bulk inversion-asymmetry (BIA) term, $H_{\textrm{BIA}}$ \cite{Ruan2016a, Dai2008}:
\begin{subequations}
    \begin{eqnarray}
    &&H_{0}(\mathbf{k}) = h_0 + \sum_{i=1}^{5}h_{i}\Gamma_i,\\
    &&H_{\textrm{BIA}}(\mathbf{k}) = \frac{2}{\sqrt{3}} C\left(k_{x} V_{x}+k_{y} V_{y}+k_{z} V_{z}\right).
    \end{eqnarray}
    \end{subequations}
Here, $h_0=E_v-\beta_c\gamma_1k^2$, $h_1=\beta_c\gamma_2\left(2k^2_z-k^2_{\|}\right)$, $h_2=\sqrt3 \beta_c\gamma_2K^2$, $h_3=2\sqrt3\beta_c\gamma_3k_xk_y$, $h_4=2\sqrt3\beta_c\gamma_3k_xk_z$, $h_5=2\sqrt3\beta_c\gamma_3k_yk_z$, $\Gamma_{1}=\frac{1}{3}\left(2 J_{z}^{2}-J_{x}^{2}-J_{y}^{2}\right)$, $\Gamma_2=\frac{1}{\sqrt{3}}\left(J_{x}^{2}-J_{y}^{2}\right)$, $\Gamma_3=\frac{2}{\sqrt{3}} J_{x y}$, $\Gamma_4=\frac{2}{\sqrt{3}} J_{z x}$, and $\Gamma_5=\frac{2}{\sqrt{3}} J_{y z}$. $J_i$ $(i=x,y,z)$ denotes spin-3/2 matrices (see the explicit matrix form in Appendix II), $J_{ij}=\frac{1}{2}\left\{J_i,J_j\right\}$, $V_{x}=\frac{1}{2}\left\{J_x,J_y^2-J_z^2\right\}$, $V_{y}=\frac{1}{2}\left\{J_y,J_z^2-J_x^2\right\}$, $V_{z}=\frac{1}{2}\left\{J_z,J_x^2-J_y^2\right\}$, and $\beta_c=\hbar^2/\left(2m'\right)$; $m'$ is the effective mass of $\Gamma_6$ bands near the $\Gamma$ point, $k^2=k_x^2+k_y^2+k_z^2$, $k_{\|}^2=k_x^2+k_y^2$, $K^2=k_x^2-k_y^2$, and \{ \} represents the anti-commutator. The AFM order is treated as a perturbation, and in the basis of $\Gamma_8$ bands, is given by \cite{Yu2017}
\begin{equation}
H_{\textrm{AFM}}(\mathbf{M}) = \xi_0 + \sum_{i=1}^{5}\xi_i\Gamma_{i},
\end{equation}
where the explicit expressions for the $\xi_i$ parameters are as given in Appendix II. Based on the low-energy effective Hamiltonian, the band structures and the phase diagrams under different directions of AFM moments can be easily obtained \cite{Yu2017}. Here, we start from the nonmagnetic phase, then focus on the typical cases with lowered symmetries, where the magnetic moments are in the $C_{3v}$ axis, parallel to mirror planes, and in a general direction, respectively.

In the nonmagnetic parent phase without AFM order, a pair of symmetry-protected TPs are present along each $C_{3v}$ axis, as exemplified by the band structure along the $[111]$ direction, as shown in Fig.~\ref{fig:phase}(a), with detailed parameters provided in Table~S1 of Appendix I.  Two TPs are formed by a doubly-degenerate downward $\Lambda_6$ band, and a nondegenerate upward $\Lambda_{5}$ $(\Lambda_4)$ band, belonging to two-dimensional and one-dimensional irreducible representations of the $C_{3v}$ point group, respectively.

\begin{figure}[htbp]
  \includegraphics[width =3.3in]{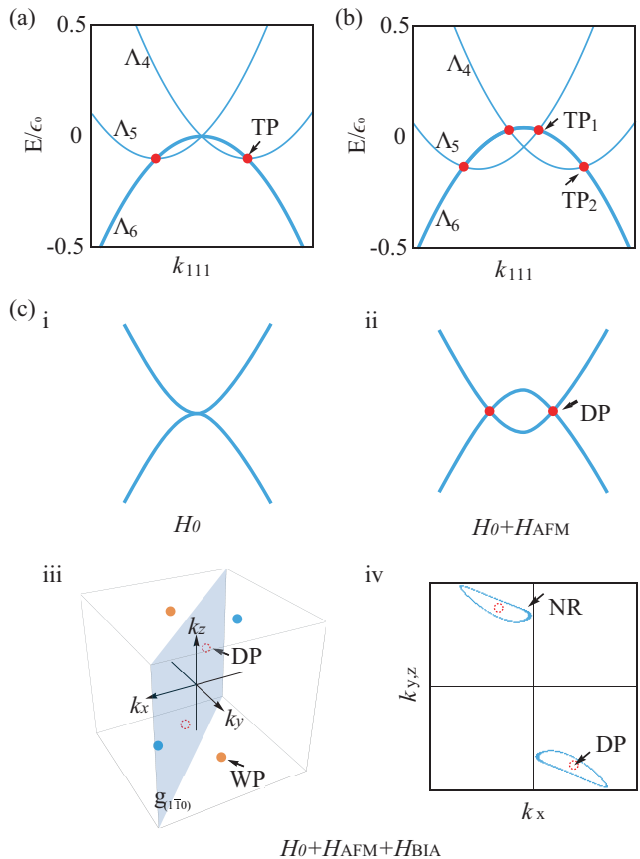}
  \caption{\label{fig:phase}(a) Band structure along the [111] direction for nonmagnetic half-Heusler materials, based on the $H_0+H_{\textrm{BIA}}$ model. Here, $\epsilon_{0}$ is chosen as the energy unit (see Table S2 in Appendix I). A pair of $C_{3v}$ symmetry-protected TPs marked by red points. (b) Band structure along the $[111]$ direction for AFM half-Heusler materials with magnetic moments in the $[111]$ direction. Note that the $C_{3v}$ symmetry is preserved. The AFM perturbation only pulls away the $\Lambda_6$ and $\Lambda_{4,5}$ bands, generating one more pair of TPs, also marked by red points. (c) Weyl semimetal and nodal-line semimetal phases of AFM half-Heusler materials with magnetic moments in $[110]$ and $[100]$ directions. (i) The schematic band structure of the $H_0$ model. A schematic of the band structure of the Dirac semimetal phase is given in (ii), based on the $H_0+H_{\textrm{AFM}}$ model. DPs are marked by red filled circles. When the BIA term is further introduced, each DP splits into a pair of WPs, or evolute into an NR, depending on the magnetic moment's direction. The Weyl semimetal phase is given in (iii), with the AFM magnetic moments in the $[110]$ direction. The blue and orange colors denote the chirality of WPs. The nodal-line semimetal phase is given in (iv), with the AFM magnetic moments in the $[100]$ direction. The red dashed circles in (iii) and (iv) represent the DPs in (ii).}
\end{figure}

When the AFM magnetic moments occur along the high-symmetry $C_{3v}$ axis (e.g., $[111]$ direction), the $C_{3,[111]}$ rotation symmetry is preserved, while all mirror symmetries in $C_{3v}$ are explicitly broken; due to the fractional translation, they are instead replaced by the glide symmetries. Nevertheless, the point group is still $C_{3v}$; as thus, the degeneracy of each band remains the same. In fact, we find that the AFM perturbation only shifts and pulls away the $\Lambda_6$ and $\Lambda_{4,5}$ bands, as shown in Fig.~\ref{fig:phase}(b) (detailed parameters are listed in Table~S2 of Appendix I), leading to the breaking of the fourfold generacy at the $\Gamma$ point and generating one more pair of TPs. This is consistent with the previous calculations \cite{Yang2017}. These TPs belong to type B, which is characterized by the presence of more than one nodal line connecting them, with the $\pi$ Berry phase of a loop encircling each nodal line \cite{Zhu2016}. 

When the magnetic moments are applied within a mirror plane, i.e., $\alpha$ [$\alpha=(1\overline10)$, $(10\overline1)$, or $(01\overline1)$], but away from the $\left[111\right]$ direction, e.g., along $\left[110\right]$, only the glide symmetry, $g_{\alpha}$, is preserved, and other symmetries are broken. In this case, the TPs along the $C_{3v}$ axes in the nonmagnetic phase are gapped or split into WPs. To illustrate this, in Fig.~\ref{fig:phase}(c)iii, we choose representative parameters (see Table~S3 in Appendix I), and schematically plot the locations of the WPs. This can be understood as follows: If we neglect the inversion-symmetry breaking BIA term, the combination of the inversion symmetry and the $\mathcal{S}$ symmetry will lead to double degeneracy for each band. The energy of the total Hamiltonian $H_0\left(\mathbf{k}\right)+H_{\textrm{AFM}}$ can be solved as follows:
\begin{equation}
    E_{\pm}(\mathbf{k})=\left(h_{0}+\xi_{0}\right) \pm \sqrt{\sum_{i=1}^{5}\left(h_{i}+\xi_{i}\right)^{2}}.
    \end{equation}
A Dirac semimetal phase is expected when $h_i+\xi_i=0$ for all $i=1,2,...,5$, as schematically shown in Fig.~\ref{fig:phase}(c)ii. The phase diagram of the Dirac semimetal phase as a function of $\xi_3/\epsilon_0$ and $\xi_4/\epsilon_0$ turns out to be a critical line \cite{Yu2017}. When including the inversion-symmetry breaking BIA term, each Dirac point (DP) will split into WPs owing to the lifting of Kramer's degeneracy, as schematically shown in Fig.~\ref{fig:phase}(c)iii. 

When the magnetic moments occur along a generic direction, e.g., the $[100]$ direction, all symmetries of $C_{3v}$ are broken. Similarly, without the inversion-breaking BIA term, the system exhibits a Dirac semimetal phase when all the $h_i+\xi_i=0$ for $i=1,2,...,5$ (see Table~S4 in Appendix I), and where the two DPs are located at the $k_y=k_z$ plane. Once the inversion-breaking BIA term is included, the DPs evolve into two nodal rings (NRs), schematically shown in Fig.~\ref{fig:phase}(c)iv, leading to a nodal-line semimetal phase.


\textit{Berry Curvature Dipole.} Firstly, we provide a brief overview of the NLHE. For a time-reversal-invariant system, an oscillating electric field $E(t)=\mathrm{Re}\{\mathcal{E}e^{i\omega t}\}$ will induce a second-order nonlinear response current, $j_a^{2\omega}=\chi_{abc}\mathcal{E}_b\mathcal{E}_c$, with the response coefficient, $\chi_{\mathrm{abc}}$, given by \cite{Sodemann2015}
\begin{equation}
\chi_{\mathrm{abc}}=-\varepsilon_{a d c} \frac{e^{3} \tau}{2(1+i \omega \tau)}D_{bd},
 \label{eq:chi}
\end{equation}
where $\varepsilon_{adc}$ is the third-rank Levi-Civita tensor, $\tau$ is the relaxation time, and
\begin{equation}
\label{Dbd}
D_{bd}=\int f_{0}(\partial_{b} \Omega_{d}) \mathbf{dk}
\end{equation}
is the dipole moment of the Berry curvature $\Omega_{d}$ over the occupied states, referred to as BCD. Here, $f_{0}$ represents the equilibrium Fermi-Dirac distribution. Note that since the BCD density, $d_{bd}(\mathbf{k})\equiv \partial_{b} \Omega_{d}(\mathbf{k})$, is odd under inversion symmetry, $D_{bd}$ vanishes in the presence of inversion symmetry, leading to the absence of a nonlinear Hall response. In contrast, $d_{bd}(\mathbf{k})$ is even under the time-reversal symmetry, so that $D_{bd}$ and corresponding NLHE persist in time-reversal-invariant systems with the breaking of inversion symmetry. It is worth mentioning that a requirement of the time-reversal symmetry is to exclude the contributions to the nonlinear Hall current from semiclassical terms which have nothing to do with the BCD \cite{Sodemann2015}.

Intriguingly, for AFM materials with broken time-reversal symmetry, as long as they possess the $\mathcal{S}$ symmetry, a combination of time-reversal symmetry and fractional translation, the BCD density $d_{bd}(\mathbf{k})$ satisfies $\mathcal{S} d_{bd}(\mathbf{k})=d_{bd}(-\mathbf{k})$. This means that $\mathcal{S}$ symmetry plays the same role as time-reversal symmetry for the BCD, such that Eq.~(\ref{eq:chi}) still holds for these AFM materials~\cite{Shao2020}. Taking into account that such $\mathcal{S}$ symmetry exists in the noncentrosymmetric AFM half-Heusler materials, one expects to observe BCD-origin NLHEs.

Point-group symmetries impose further constraints on the BCD tensor, $D_{bd}$, and force some components to vanish. In Table~\ref{table:symmetries}, we list the transformation properties of $k_{i}$ and $\Omega_{i}$, where $i=x,y,z$, under several important symmetry operations of the $C_{3v}$ group in the Cartesian coordinate for AFM half-Heusler materials. According to Table~\ref{table:symmetries}, when the magnetic moments are in the $[111]$ direction, preserving the $C_{3v}$ group, we find that $D_{xy}=-D_{yx}=D_{yz}=-D_{zy}=D_{zx}=-D_{xz}$ and the rest components vanish, because their BCD density is odd under at least one of the symmetry operations. This means that only one of the nine components of $D_{bd}$ is independent in this case. When the magnetic moments are in the $[110]$ direction, only one glide symmetry $g_{(1\overline10)}$ is preserved. According to Table~\ref{table:symmetries}, we obtain $D_{xx}=-D_{yy}$, $D_{zz}=0$, $D_{xy}=-D_{yx}$, $D_{zx}=-D_{zy}$, and $D_{xz}=-D_{yz}$, resulting in four independent components. If the magnetic moments are in the $[100]$ direction, which breaks all symmetries in $C_{3v}$ group, all nine BCD components persist, and are generally independent from one another. Moreover, for the nonmagnetic phase belonging to the $T_d$ group, symmetries force all BCD components to vanish in the same way \cite{Chen2019}. As a result, depending on the directions of the AFM moments, different nonvanishing and independent BCD components are expected, due to the gradually lowered symmetry, as summarized in Table~\ref{table:numbers}. Therefore, the nonzero BCD components obtained via experimental NLHE measurements can be used to determine whether the AFM moments are in a high-symmetry axis, within high symmetry planes, or in a general direction. It should be emphasized that the above symmetry arguments are quite general, and can be applied to a wide class of AFM Half-Heusler materials.

\begin{table}[htbp]
    \centering
    \caption{\label{table:symmetries}Properties of $k_{i}$ and $\Omega_{i}$ under several important symmetry operations in the $C_{3v}$ group.}
    \setlength{\tabcolsep}{3.5mm}
    \begin{tabular} {|c|c|c|c|c|c|c|}
      \hline
       & $k_{x}$ & $k_{y}$  & $k_{z}$ & $\Omega_{x}$ & $\Omega_{y}$ & $\Omega_{z}$\\
      \hline
       $g_{(1\overline10)}$ & $ k_y $ & $ k_x $ &$k_z$ &$-\Omega_y$ &$-\Omega_x$ & $-\Omega_z$\\
      \hline
        $g_{(10\overline1)}$& $k_z$ & $k_y$ &$k_x$ &$-\Omega_z$ &$-\Omega_y$ & $-\Omega_x$\\
      \hline
        $g_{(01\overline1)}$ & $k_x$ & $k_z$ & $k_y$ & $-\Omega_x$ & $-\Omega_z$ & $-\Omega_y$ \\
      \hline
        $C_{3,[111]}$ & $k_z$ & $k_x$ & $k_y$ & $\Omega_z$ & $\Omega_x$ & $\Omega_y$ \\
      \hline
    \end{tabular}
\end{table}

\begin{table}[htbp]
    \centering
    \caption{\label{table:numbers} Numbers of nonzero BCD components and independent BCD components for different magnetic orders.}
    \setlength{\tabcolsep}{3.5mm}
    \begin{tabular} {|c|c|c|}
      \hline
      Magnetic orders & Nonzero & Independent \\
      \hline
       Nonmagnetic & 0 & 0 \\
      \hline
        $\mathbf{M}_{[111]}$ & 6 & 1 \\
      \hline
        $\mathbf{M}_{[110]}$ & 8 & 4  \\
      \hline
        $\mathbf{M}_{[100]}$ & 9 & 9  \\
      \hline
    \end{tabular}
\end{table}

Next, we focus on the TP semimetal and Weyl semimetal phases of AFM half-Heusler materials under magnetic moments in the $[111]$ and $[110]$ directions, respectively, so as to explicitly calculate the BCD as a function of energy. According to Eq.~(\ref{Dbd}), two factors, namely, the Fermi distribution $f_{0}$ and the BCD density $d_{bd}(\mathbf{k})$, are involved in the calculation of $D_{bd}$. For simplicity, the temperature is set to be zero such that $f_{0}(\mathbf{k})=1$/$0$ for occupied/unoccupied states. In addition, the Berry curvature of the $i$th band can be obtained via the following formula~\cite{Xiao2010}:
\begin{equation}
        \Omega_{i \mathbf{k}}^{a}=-2 \varepsilon^{a b c} \sum_{j \neq i}
        \frac{\operatorname{Im}\left\langle i\left|\partial_{k_{b}} \hat{\mathcal{H}}
        \right| j\right\rangle\left\langle j\left|\partial_{k_{c}}
        \hat{\mathcal{H}}\right| i\right\rangle}
        {\left(\epsilon_{\mathbf{k}}^{i}-\epsilon_{\mathbf{k}}^{j}\right)^{2}},
\end{equation}
where $a,b,c=x,y,z$, and $i,j$ are band indices. Here, the low-energy physics of the AFM half-Heusler materials can be properly described by the effective $k\cdot p$ Luttinger model around $\Gamma$ point, so the BCD can be simply calculated by the effective model, without involving the full Brillouin zone. Taking $D_{xy}$ as an example, which corresponds to $j_x^{2\omega}=\chi_{xxz}\mathcal{E}_x\mathcal{E}_z$, our results are shown in Fig.~\ref{fig:BCD}, where the parameters for the calculations are listed in Table~S2 and Table~S3 in Appendix I. A prominent peak of BCD around WPs is found in Weyl semimetal phase, while there is no pronounced contribution from TPs in TP semimetal phase. There are some weak peaks around the TPs, which are mainly contributed by other trivial Fermi pockets. 

\begin{figure}[htbp]
    \includegraphics[width =3.4in]{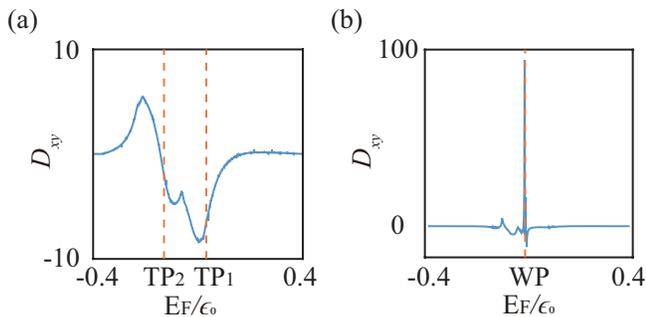}
    \caption{\label{fig:BCD} Calculated BCD component, $D_{xy}$, with varying Fermi energy, for (a) TP semimetal and (b) Weyl semimetal phases of the AFM half-Heusler materials under $[111]$- and $[110]$-direction magnetic moments, respectively. A prominent peak of BCD around the WPs is found in Weyl semimetal phase [see the red dashed line in (b)], while there is no pronounced contribution from TPs in TP semimetal phase.}
\end{figure}

We further analyze the contributions of TPs and WPs based on the local general Hamiltonian near these band crossing points. The contributions of WPs have been discussed in previous works \cite{Sodemann2015, Facio2018, Zhang2018c, Chen2019}, which point out that magnitude of the BCD can be tuned by the tilt of the Weyl cone. The type-B TPs can be described by the low-energy effective Hamiltonian expanded to linear order around the TPs \cite{Winkler2016}:
\begin{small}
\begin{equation}
    H=\left(
        \begin{array}{cccc}
            E_{0}+A k_{z} & 0 & D k_{y} & D k_{x} \\
            0 & -E_{0}+A k_{z} & F^{*} k_{x} & -F^{*} k_{y} \\
            D^{*} k_{y} & F k_{x} & B k_{z}+C k_{x} & C k_{y} \\
            D^{*} k_{x} & -F k_{y} & C k_{y} & B k_{z}-C k_{x}
        \end{array}
        \right).
\end{equation}
\end{small}
Here, $A, B, C, D (D^{*}), E_{0},  F (F^{*})$ are parameters, and the $C_{3v}$ direction $\left[111\right]$ is chosen as the $k_z$ axis. For the doubly degenerate bands belong to the two-dimensional representation in the $k_z$ direction, the low-energy Hamiltonian around each TP in the $k_x-k_y$ plane to linear order is given by
\begin{equation}
    H=B k_{z} \sigma_{0}+C k_{x} \sigma_{z}+C k_{y} \sigma_{x},
    \end{equation}
which clearly leads to vanishing Berry curvature. For the two nondegenerate bands, the low-energy Hamiltonian to linear order is given by
\begin{equation}
    H=A k_{z} \sigma_{0}+E_{0} \sigma_{z},
    \end{equation}
which also gives rise to zero Berry curvature. Therefore, we would expect the TPs to make no contribution to BCD, which explains the numerical results above.


In summary, we have investigated the NLHE in AFM half-Heusler materials. Based on the effective model method, we have systematically studied their electronic structure and phase diagrams. Due to the constraints on the BCD tensor by symmetries, some elements of BCD tensor survive for AFM half-Heusler materials. A prominent peak of BCD around WPs is found in Weyl semimetal phase, while there are no pronounced contributions from TPs in TP semimetal phase. This is further supported by an analysis based on low-energy effective Hamiltonians. Our work shows that the NLHE method may provide a powerful tool to characterize magnetic topological states with various AFM configurations in AFM half-Heusler materials.

\begin{acknowledgments}
This work was supported by the National Natural Science Foundation of China (No. 11834006, No. 12074181 and No. 11674165), the Natural Science Foundation of Jiangsu Province (No. BK20200007), the Fok Ying-Tong Education Foundation of China (Grant No. 161006), and the Fundamental Research Funds for the Central Universities (No. 020414380149).
\end{acknowledgments}

\bibliography{NLHE.bib}

\end{document}